\documentclass[11pt,a4paper]{article}
\pdfoutput=1

\usepackage[nosort]{cite}

\usepackage[british]{babel} 

\usepackage{amsmath,amssymb,amsthm}
\usepackage{mathrsfs}
\usepackage{bbm}
\usepackage{bm}
\usepackage{slashed}
\usepackage{braket}
\usepackage{dsfont}

\usepackage{color}
\usepackage[textwidth = 430 pt, textheight = 630 pt]{geometry}
\definecolor{MyDarkBlue}{rgb}{0.15,0.25,0.45}

\usepackage[linktocpage=true,hypertexnames=false]{hyperref}
\hypersetup{
colorlinks=true,
citecolor=MyDarkBlue,
linkcolor=MyDarkBlue,
urlcolor=MyDarkBlue,
pdfauthor={Andrew Rolph${}^a$ and Alessandro Torrielli${}^b$},
pdftitle={Secret Symmetries of Type IIB Superstring Theory on AdS3 x S3 x M4},
pdfsubject={hep-th}
breaklinks=true
}

\let\fn\footnote
\renewcommand{\footnote}[1]{\linespread{1.1}\fn{#1}\linespread{1.29}}

\makeatletter\renewcommand{\section}{\@startsection
{section}{1}{\z@}{-3.5ex plus -1ex minus
    -.2ex}{2.3ex plus .2ex}{\bf\mathversion{bold} }}
\makeatletter\renewcommand{\subsection}{\@startsection{subsection}{2}{\z@}{-3.25ex
plus -1ex minus
   -.2ex}{1.5ex plus .2ex}{\bf\mathversion{bold} }}
\makeatletter\renewcommand{\subsubsection}{\@startsection{subsubsection}{3}{-2.45ex}{-3.25ex
plus -1ex minus -.2ex}{1.5ex plus .2ex}{\it }}
\renewcommand{\thesection}{\arabic{section}}
\renewcommand{\thesubsection}{\arabic{section}.\arabic{subsection}}
\renewcommand{\@seccntformat}[1]{\@nameuse{the#1}.~~}

\renewcommand{\theequation}{\thesection.\arabic{equation}}
\makeatletter \@addtoreset{equation}{section}

\renewcommand*\l@section{\@dottedtocline{1}{0em}{2em}}
\renewcommand*\l@subsection{\@dottedtocline{2}{2em}{2.4em}}
\renewcommand*\l@subsubsection{\@dottedtocline{4}{3.8em}{3.7em}}

\renewcommand\tableofcontents{%
    \section*{\large\contentsname
        \@mkboth{%
          \MakeUppercase\contentsname}{\MakeUppercase\contentsname}}%
       {\baselineskip=15pt plus 2pt minus 1pt
    \@starttoc{toc}}%
}

\renewenvironment{thebibliography}[1]
     {\baselineskip=16pt plus 2pt minus 1pt
      \section*{\large\refname
        \@mkboth{\MakeUppercase\refname}{\MakeUppercase\refname}}%
     \list{\@biblabel{\@arabic\c@enumiv}}%
           {\settowidth\labelwidth{\@biblabel{#1}}%
            \leftmargin\labelwidth
            \advance\leftmargin\labelsep
            \@openbib@code
            \usecounter{enumiv}%
            \let\p@enumiv\@empty
            \renewcommand\theenumiv{\@arabic\c@enumiv}}%
      \sloppy
      \clubpenalty4000
      \@clubpenalty \clubpenalty
      \widowpenalty4000%
      \sfcode`\.\@m
 \catcode`\^^M=10%
}

\newcommand{\appendices}{
\section*{Appendix}\label{appendices}\setcounter{subsection}{0}
\addcontentsline{toc}{section}{Appendix}
\setcounter{equation}{0}
\makeatletter
\renewcommand{\theequation}{\Alph{subsection}.\arabic{equation}}
\renewcommand{\thesubsection}{\Alph{subsection}}
\@addtoreset{equation}{subsection}
\makeatother
}

\def\XXint#1#2#3{{\setbox0=\hbox{$#1{#2#3}{\int}$}
    \vcenter{\hbox{$#2#3$}}\kern-.5\wd0}}

\newcommand{\alg}[1]{\mathfrak{#1}}

\begin{document}

\begin{titlepage}

\setcounter{page}{0}
\renewcommand{\thefootnote}{\fnsymbol{footnote}}

\begin{flushright}
DMUS--MP--14/16
\end{flushright}

\vspace{1cm}

\begin{center}

\textbf{\Large\mathversion{bold} Drinfeld basis for string-inspired Baxter operators}

\vspace{1cm}

{\large Andrew Rolph and Alessandro Torrielli \footnote{{\it E-mail addresses:\/}
\href{mailto:andrew.d.rolph@gmail.com}{\ttfamily andrew.d.rolph@gmail.com},
\href{mailto:a.torrielli@surrey.ac.uk}{\ttfamily a.torrielli@surrey.ac.uk}}
} 

\vspace{1cm}

\it Department of Mathematics, University of Surrey\\
Guildford GU2 7XH, United Kingdom

\vspace{1cm}

{\bf Abstract}
\end{center}
\vspace{-.3cm}
\begin{quote}
We propose Drinfeld's second realisation of the quantum group relevant to the Lax-operator approach developed in the work of  Bazhanov, Frassek, Lukowski, Meneghelli and Staudacher.

\vfill
\noindent December 2014

\end{quote}

\setcounter{footnote}{0}\renewcommand{\thefootnote}{\arabic{thefootnote}}

\end{titlepage}

\tableofcontents

\bigskip
\bigskip
\hrule
\bigskip
\bigskip

\section{Introduction}

In recent years, tremendous progress has been made towards a complete solution of the AdS/CFT integrable system \cite{Reviews}, with the two sides of the correspondence matching to a dazzling degree of sophistication and accuracy. Nevertheless, there remains a feeling that a full mathematical understanding of the mechanism may require additional investigation. One route to gather more data which is currently being pursued is the exploration of similar integrable structures in lower-dimensional instances of the correspondence \cite{LowerD}.  

\smallskip

In \cite{BFLMS}, on the other hand, a program was started aimed at investigating the algebraic properties of the original AdS${}_5$ system by means of the Baxter $Q$-operator approach. This program began with a revisitation of the spin-$\frac{1}{2}$ Heisenberg spin-chain in the light of $Q$-operators, where interesting `elementary' (or {\it partonic}) Lax operators were employed, and their algebraic relations and representation theory studied. These objects were then related to the theory of Yangians, hence it is natural to ask whether one might develop an analogue of Drinfeld's second realisation \cite{DII} for their description.   

\smallskip

In this brief note, we propose an answer to this question. We derive a set of defining relations for the analogue of the Drinfeld generators of ordinary Yangians, by mimicking the procedure of triangular factorisation that works in the standard case. We obtain the relevant oscillator representation which makes contact with \cite{BFLMS}, and construct the Hopf algebra maps which turn our structure into a quantum group. The resulting formulae bear resemblance to the ordinary Yangian, but also present some more uncommon features which make for an interesting algebraic object.    

\smallskip

We should mention work that relates to ours, and which represents an important concurrent line of investigation in this direction. The partonic Lax operator originally appeared in the literature in connection with the so-called ``Discrete Self-Trapping" (DST) chain \cite{KuznetsovSalernoSklyaninKovalskyPronko}. The reader is referred to \cite{Zengo} for further details\footnote{The authors thank Zengo Tsuboi for very useful comments and guidance to the relevant literature.}. Furthermore, in the work of Chicherin, Derkachov, Karakhanyan and Kirschner \cite{Chicherin} general solutions for the Baxter operators relevant to our discussion are obtained, and similar factorisation properties to the ones we will use here are employed.

\section{The $\alg{sl}(2)$ quantum group}

\subsection{The standard case}

A standard and all-important object in the theory of Yangians is represented by the following rational R-matrix:
\begin{eqnarray}
\label{R}
R(u) \, = \, u \, \mathbbmss{1} + \, {\cal P},
\end{eqnarray}
with $\cal{P}$ the permutation operator. Starting from the R-matrix restricted to the $\alg{sl}(2)$ Lie algebra, one considers the Lax operator
\begin{eqnarray}
\label{firstLax}
L(z) \, = \, \begin{pmatrix}
z + h&f\\
e&z-h
\end{pmatrix},
\end{eqnarray}
with the generators appearing satisfying the $\alg{sl}(2)$ commutation relations 
\begin{eqnarray}
[h,e] \, = \, e, \qquad [h,f] \, = \, - f, \qquad [e,f] \, = \, 2 h.
\end{eqnarray}   
One can check that this is equivalent to   
\begin{eqnarray}
\label{RTT}
R(u-v) \, [L(u) \otimes \mathbbmss{1}] \, [\mathbbmss{1} \otimes L(v)] \, = \, [\mathbbmss{1} \otimes L(v)]  \, [L(u) \otimes \mathbbmss{1}] \, R(u-v),
\end{eqnarray}
or, in other words, to
\begin{eqnarray}
\label{into}
(u-v) \, [L_{ij}(u), L_{hk} (v)] \, = \, L_{hj}(u) \, L_{ik} (v) \, - \, L_{hj}(v) \, L_{ik} (u),
\end{eqnarray}
for $i,j,h,k=1,2$. This provides the starting point for the quantum inverse scattering method \cite{Evgeny}.

\smallskip 

\subsection{Gauss decomposition}

The next step towards Drinfeld's second realisation of the ordinary Yangian is to apply the following {\it Gauss decomposition} \cite{BC,BCC,BC2}:
\begin{eqnarray}
\label{as}
L(z) \, = \, \begin{pmatrix}
1&0\\
F(z)&1
\end{pmatrix}\begin{pmatrix}
D_1(z)&0\\
0&D_2(z)
\end{pmatrix}\begin{pmatrix}
1&E(z)\\
0&1
\end{pmatrix} \, = \, \begin{pmatrix}
D_1&D_1 \, E\\
F \, D_1&F \, D_1 \, E \, + \, D_2
\end{pmatrix}.
\end{eqnarray}
One can determine the commutation relations for the {\it currents} $D_i(z)$, $E(z)$, $F(z)$ by substituting the above decomposition (\ref{as}) into (\ref{into}), or equivalently into (\ref{RTT}) using (\ref{R}). This
gives:

\begin{eqnarray}
\label{among}
&&[D_i(u), D_j(v)] \, = \, 0, \qquad i,j=1,2\nonumber\\
&&(u-v) \, [D_1(u), E(v)] \, = \, D_1(u) [E(v) - E(u)],\nonumber\\
&&(u-v) \, [D_1(u), F(v)] \, = \,  [F(u) - F(v)]D_1(u),\nonumber\\
&&(u-v) \, [D_2(u), E(v)] \, = \,  D_2(u)[E(u) - E(v)],\nonumber\\
&&(u-v) \, [D_2(u), F(v)] \, = \,  [F(v) - F(u)]D_2(u),\nonumber\\
&&(u-v) \, [E(u), E(v)] \, = \,  [E(u) - E(v)]^2,\nonumber\\
&&(u-v) \, [F(u), F(v)] \, = \, -[F(u) - F(v)]^2,\nonumber\\
&&(u-v) \, [E(u), F(v)] \, = \, D_1^{-1}(u) D_2(u) - D_1^{-1}(v) D_2(v).
\end{eqnarray}

Drinfeld's second realisation of the ordinary Yangian is at this point typically achieved by expanding the above currents in {\it modes} (see below), and deducing how the relations (\ref{among}) translate to relations on the modes of the expansion. We will illustrate this procedure in the next subsection for the novel case discussed in this paper. 
 
\smallskip

\subsection{The novel case}

The authors of \cite{BFLMS} consider instead an alternative ({\it partonic}) Lax operator:
\begin{eqnarray}
\label{BFLMS}
L(z) \, = \, \begin{pmatrix}
z + h&a^\dagger\\
a&1
\end{pmatrix},
\end{eqnarray}
with
\begin{eqnarray}
\label{Heise}
[a, a^\dagger]=1, \qquad h=a^\dagger \, a.
\end{eqnarray}   
One can check that (\ref{RTT}) still holds.

\smallskip

Our idea is to adopt the very same decomposition as in (\ref{as}) for this new Lax operator. When we then come to expressing the currents in terms of modes, in order to obtain a natural identification with the form of (\ref{BFLMS}), we find it necessary to impose a slightly different expansion then the one used for standard Yangians in \cite{BCC}. We employ the following mode-expansion\footnote{The change with respect to \cite{BCC} is only in the Cartan part $\Gamma(u)$.}:
\begin{eqnarray}
\label{exp}
E(u) = \sum_{k=0}^\infty \, \xi^+_k \, u^{-k-1}, \qquad F(u) = \sum_{k=0}^\infty \, \xi^-_k \, u^{-k-1}, \qquad \Gamma (u) = 1 - \sum_{k=0}^\infty \, \kappa_k \, u^{-k-1}, 
\end{eqnarray}
with
\begin{eqnarray}
\Gamma (u) = 1 + D_1^{-1}(u) D_2(u).
\end{eqnarray}
This implies that (\ref{among}) is equivalent to the following system of defining relations\footnote{A similar construction for the Lax operator (\ref{firstLax}) would produce the defining relations of the Yangian in Drinfeld's second realisation \cite{DII,BC,BCC}.}:
\begin{eqnarray}
\label{HII}
&&[\kappa_n, \kappa_m]=0, \qquad [\kappa_0, \xi^\pm_m] = 0,\nonumber\\
&&[\xi^+_m, \xi^-_n] = \kappa_{m+n},\nonumber\\
&&[\kappa_m, \xi^\pm_{n+1}] - [\kappa_{m+1}, \xi^\pm_n] = \pm \{ \kappa_m, \xi^\pm_n \},\nonumber\\
&&[\xi^\pm_m, \xi^\pm_{n+1}] - [\xi^\pm_{m+1}, \xi^\pm_n] = \pm \{ \xi^\pm_m, \xi^\pm_n \}.
\end{eqnarray}

These relations are different from those of the standard Yangian \cite{DII,BCC,BC2}. The crucial distinction at this stage seems to be the central element $\kappa_0$ and the relations involving it. We will see in the next section that more differences are to be observed at the co-algebra level, when we will define a coproduct compatible with (\ref{HII}). 

\medskip

The representation one obtains from the explict use of (\ref{BFLMS}) is as follows:
\begin{eqnarray}
\label{eval}
\xi^+_m = (-h)^m \, a^\dagger, \qquad \xi^-_m = a \, (-h)^m, \qquad \kappa_m = (-)^m [\, h^{m+1} - (h+1)^{m+1} \,].
\end{eqnarray}

\smallskip

Let us make here an interesting observation\footnote{We thank the referee for pointing out this possibility to us.}. Had we exchanged the rows and colums of (\ref{BFLMS}) while adopting the same decomposition (\ref{as}), we would have obtained the following:
\begin{eqnarray}
\label{manifest}
D_1(u) = 1, \qquad E(u) = a, \qquad F(u) = a^\dagger, \qquad D_2(u) = u.
\end{eqnarray}
By construction, this still satisfies the whole set of relations (\ref{among}) - most of which now reduces to $0=0$ identities. However, one can observe from (\ref{manifest}) that such an assignment extracts from our new quantum group (\ref{HII}) the ``level-zero" subalgebra only, {\it i.e.} the Heisenberg algebra (\ref{Heise}).

\smallskip

To obtain an {\it evaluation} representation depending on a spectral parameter $\lambda$ one just needs to substitute $h \rightarrow h+\lambda$ in $\xi^\pm_m$ in (\ref{eval}) (and re-calculate $\kappa_m$, for instance as $[\xi^+_0, \xi^-_m]$). In fact, the quantum group we have found is invariant under the {\it shift automorphism}
\begin{eqnarray}
\kappa_1 \, \longrightarrow \kappa_1 + \mu \, \kappa_0, \qquad \xi^\pm_1 \, \longrightarrow \xi^\pm_1 + \mu \, \xi^\pm_0,
\end{eqnarray}
with a complex parameter $\mu$.

\medskip

One thing to notice is that there is no obvious way to ``mechanically" generate all higher levels given the level $0$ and $1$ generators, as $\kappa_0$ is central. This is at odds with the case of the standard Yangian, where such a mechanical procedure can be found \cite{Levendo}. It also means that we should pay particular attention to all the statements that extend beyond level $1$.

\medskip

We can perform a map to the analogue of Drinfeld's first realisation of our quantum group. Let us define
\begin{eqnarray}
\label{hat}
\widehat{\kappa} \, \equiv \, \frac{\kappa_1}{2} \, - \, \xi^-_0 \, \xi^+_0, \qquad \widehat{\xi}^+ \, \equiv \, \xi^+_1 \, - \, \frac{1}{2} \, \kappa_0^{-1} \, \kappa_1 \, \xi^+_0, \qquad \widehat{\xi}^- \, \equiv \, \xi^-_1 \, - \, \frac{1}{2} \, \kappa_0^{-1} \,  \xi^-_0 \, \kappa_1,
\end{eqnarray}
where we have allowed ourselves to take inverse powers of the central element $\kappa_0$.
One can show that the new generators satisfy $[T^a,\widehat{T}^b]=f^{ab}_c \,\widehat{T}^c$, namely
\begin{eqnarray}
[\kappa_0, \, \, \widehat{.}\, \, ] \, = \, 0 \, = [\xi^\pm_0, \widehat{\kappa}]\, \qquad \qquad [\xi^\mp, \widehat{\xi}^\pm_0] \, = \, \mp \, \widehat{\kappa},
\end{eqnarray}
where $\widehat{.}$ denotes any of the generators (\ref{hat}).
\medskip

\subsection{Coproduct}

The coproduct we can equip our algebra with is rather uncommon. It can be obtained from the familiar formula
\begin{eqnarray}
\Delta\big(L_{ij}(u)\big) \, = \, L_{ik}(u) \otimes L_{kj}(u),
\end{eqnarray}
however the expansion in powers of $u$ reserves a few surprises with respect to the analogous computation performed for the ordinary Yangian. 
Expanding as in (\ref{exp}), and with a few manipulations, we obtain, for example for the first two levels, the unusual
\begin{eqnarray}
\label{copo}
&&\Delta(\xi^+_0) \, = \, 1\otimes \xi^+_0, \qquad \Delta(\xi^+_1) \, = \, 1\otimes \xi^+_1 + \xi^+_0 \otimes \kappa_0,\nonumber\\
&&\Delta(\xi^-_0) \, = \, \xi^-_0 \otimes 1, \qquad \Delta(\xi^-_1) \, = \, \xi^-_1\otimes 1 + \kappa_0 \otimes \xi^-_0,\nonumber\\
&&\Delta(\kappa_0) \, = \, 0, \qquad \Delta(\kappa_1) \, = \, \kappa_0\otimes  \kappa_0.
\end{eqnarray}
One can check that this satisfies (\ref{HII}), providing a homomorphism of the set of definining relations.

The coproduct (\ref{copo}) is rather dissimilar from the one typically assigned to the standard Yangian \cite{DII,BCC,BC2}. The main diffence is in the ``level-zero" comultiplication rule, which is surprisingly {\it a-symmetric}. Where one would have expected a formula of the type $\Delta(\alg{t}_0) = \alg{t}_0 \otimes 1 + 1 \otimes \alg{t}_0$, one sees instead a quite peculiar ``halved" coproduct in (\ref{copo})\footnote{One could almost attempt to say that the partonic Lax operator of \cite{BFLMS} produces a {\it partonic} coproduct, meaning that the comultiplication map itself gets also decomposed into some more elementary building blocks (its two ``halves").}. We believe this to be the most significant piece of contradistinction of our new quantum group with respect to the ordinary Yangian.

\section{Conclusions}

In this short letter, we have proposed Drinfeld's second realisation for the Baxter-operator approach of \cite{BFLMS}. This realisation is traditionally best suited to derive important algebraic quantities, such as the universal R-matrix, and to study representations of the ordinary Yangian. We believe therefore that the realisation we have derived might play the same role in the present case, and favour progress in the program inaugurated by the authors of \cite{BFLMS}.

\smallskip

We have found that the defining relations and Hopf-algebra coproduct which characterise our quantum group are somewhat similar to the standard ones, however they present some crucial differences which stimulate curiosity in this novel algebraic structure. We plan to come back to the full investigation of these aspects in future work.

\section{Acknowledgements}

AT thanks the EPSRC for funding under the First Grant project EP/K014412/1 ``Exotic quantum groups, Lie superalgebras and integrable systems", and the STFC for support under the Consolidated Grant project nr. ST/L000490/1 ``Fundamental Implications of Fields, Strings and Gravity". AT acknowledges useful conversations with the participants of the ESF and STFC supported workshop ``Permutations and Gauge String duality" (STFC-4070083442, Queen Mary U. of London, July 2014). AT thanks Vladimir Kazakov for very useful discussions, and Zengo Tsuboi for feedback after the presentation of some preliminary results in a group seminar at the Humboldt University of Berlin, and for a consequent very helpful email exchange. AT is also indebted to both theory groups of the Humboldt University, and in particular to Harald Dorn, Tomek Lukowski, Vladimir Mitev, Jan Plefka and Matthias Staudacher, for kind hospitality and stimulating discussions on that occasion.

\end{document}